\documentclass[twocolumn,prx,english,superscriptaddress,longbibliography]{revtex4-2}
\usepackage[T1]{fontenc}
\usepackage[latin9]{inputenc}
\usepackage{geometry}
\usepackage[normalem]{ulem}
\usepackage{xcolor}
\usepackage{babel}
\usepackage{units}
\usepackage{braket}
\usepackage{amsmath}
\usepackage{amssymb}
\usepackage{stmaryrd}
\usepackage{graphicx}
\usepackage{wasysym}
\usepackage{bbold}
\usepackage[colorlinks=true,citecolor=magenta,linkcolor =blue]{hyperref}
\usepackage[all]{hypcap} 
\usepackage{times}
\usepackage{orcidlink}
\usepackage{lineno}
\usepackage[normalem]{ulem}
\makeatletter

\IfFileExists{lmodern.sty}{\usepackage{lmodern}}{}
\usepackage{babel}

\usepackage{hyperref}
\makeatother

\begin{document}
\title{Hybrid digital-analog protocols for simulating quantum multi-body interactions}

\author{Or Katz}
\thanks{These authors contributed equally to this work.}
\affiliation{School of Applied and Engineering Physics, Cornell University, Ithaca, NY 14853}

\author{Alexander Schuckert}
\thanks{These authors contributed equally to this work.}
\affiliation{Joint Center for Quantum Information and Computer Science, 
        NIST/University of Maryland, College Park, MD 20742, USA}
    \affiliation{Joint Quantum Institute, NIST/University of Maryland, 
        College Park, MD 20742, USA}

\author{Tianyi Wang}
\thanks{These authors contributed equally to this work.}
\affiliation{Department of Physics, Duke University, Durham, NC 27708}
\affiliation{Duke Quantum Center, Duke University, Durham, NC 27701}

\author{Eleanor Crane}
\affiliation{Department of Physics, King's College London, Strand, London, WC2R 2LS, UK}

\author{Alexey V. Gorshkov}
\affiliation{Joint Center for Quantum Information and Computer Science, 
        NIST/University of Maryland, College Park, MD 20742, USA}
    \affiliation{Joint Quantum Institute, NIST/University of Maryland, 
        College Park, MD 20742, USA}

\author{Marko Cetina}
\affiliation{Department of Physics, Duke University, Durham, NC 27708}
\affiliation{Duke Quantum Center, Duke University, Durham, NC 27701}
\affiliation{Department of Electrical and Computer Engineering, Duke University, Durham, NC 27708}

\begin{abstract}
While quantum simulators promise to explore quantum many-body physics beyond classical computation, their capabilities are limited by the available native interactions in the hardware. On many platforms, accessible Hamiltonians are largely restricted to one- and two-body interactions, limiting access to multi-body Hamiltonians and to systems governed by simultaneous, non-commuting interaction terms that are central to condensed matter, quantum chemistry, and high-energy physics. We introduce and experimentally demonstrate a hybrid digital-analog protocol that overcomes these limitations by embedding analog evolution between shallow entangling-gate layers. This method produces effective Hamiltonians with simultaneous non-commuting three- and four-body interactions that are generated non-perturbatively and without Trotter error -- capabilities not practically attainable on near-term hardware using purely digital or purely analog schemes. We implement our scheme on a trapped-ion quantum processor and use it to realize a topological spin chain exhibiting prethermal strong zero modes persisting at high temperature, as well as  models featuring three- and four-body interactions. Our hardware-agnostic and scalable method opens new routes to realizing complex many-body physics across quantum platforms.
\end{abstract}
\maketitle

\section*{Introduction} 
\vspace{-5pt}
Quantum simulators offer a powerful route to exploring many-body dynamics beyond the reach of classical computation. While universal simulators can, in principle, approximate arbitrary dynamics, the models that can be simulated most efficiently are dictated by the native interactions in the hardware and their programmability. On most platforms, these interactions are limited to local one- and two-body terms, so realizing non-commuting higher-order ($k>2$ bodies) Hamiltonians requires circuit-depth or runtime overhead that remains beyond current experimental capabilities. Extending the set of Hamiltonians that can be natively implemented offers a pathway to simulate new classes of models that are central to condensed matter, quantum chemistry, and high-energy physics and expand the reach of quantum simulation.

Existing approaches to quantum simulation fall broadly into digital and analog paradigms, which have complementary strengths and limitations. Digital quantum simulation relies on mapping a target Hamiltonian onto sequences of native gates, typically one- and two-qubit operations, that approximate the desired time evolution~\cite{lloyd_universal_1996}. This approach is universal and, in principle, capable of simulating arbitrary dynamics. In practice, however, realizing nontrivial Hamiltonians often requires deep quantum circuits, especially when the model involves higher-order interactions or non-commuting terms. A common strategy to approximate such dynamics is Trotterization~\cite{childs_theory_2021}, which decomposes the evolution into a sequence of discrete steps under simpler sub-Hamiltonians. While formally exact in the limit of infinitely many steps, achieving high accuracy requires a polynomially large number of gates, resulting in long circuits that limit evolution to relatively short times and increase susceptibility to errors~\cite{kim_evidence_2023, hemery_measuring_2024, haghshenas_digital_2025}. This tradeoff between accuracy and circuit depth remains a key obstacle for digital simulation on near-term devices.

\begin{figure*}[htbp]
\begin{centering}
 \includegraphics[width=6.8in]{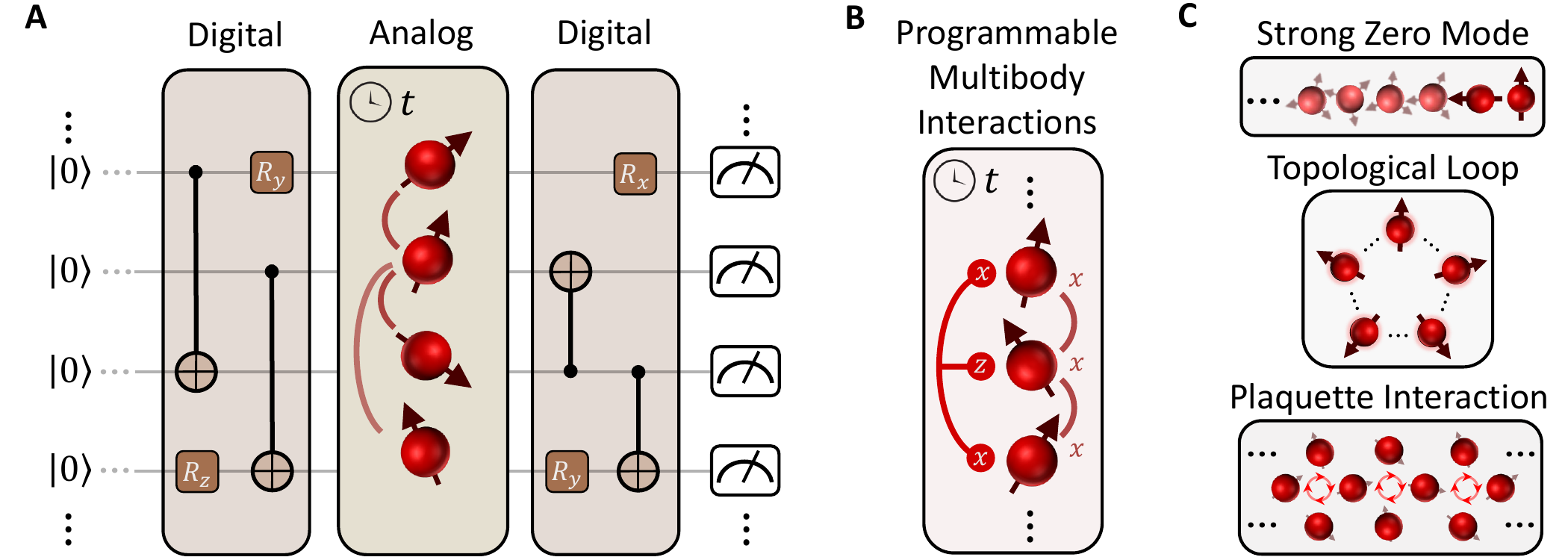}
\par\end{centering}
\centering{}\caption{\textbf{Direct multi-body interactions beyond Trotterization Concept.} \textbf{a}, Continuous analog evolution under native long-range couplings and local fields, sandwiched between two shallow digital gate layers, can generate concurrent, non-commuting multi-body interactions without Trotter discretization error. \textbf{b}, The resulting effective Hamiltonians are programmable, with tunable two-, three-, and four-body couplings, enabling simulation of diverse many-body models implemented without Trotterization. \textbf{c}, Example realizations include an open chain with a strong zero mode, a periodic topological loop, and a plaquette model with four-body interactions.\label{fig:Hero}}
\end{figure*}

Analog quantum simulation \cite{altman2021quantum} offers a complementary approach in which the system evolves under engineered interactions that directly implement a model Hamiltonian~\cite{kim_quantum_2010,trotzky_probing_2012,jurcevic_quasiparticle_2014}. This method provides distinct advantages, including the ability to apply multiple interaction terms simultaneously and to capture the dynamics of non-commuting terms without Trotterization. However, analog simulation is often less flexible than digital approaches and tends to be most effective when the desired dynamics closely match the platform's native couplings. These interactions are typically limited to specific one- or two-body terms and have restricted connectivity, which narrows the class of accessible Hamiltonians.

To overcome these limitations, strategies have been developed to extend the capabilities of quantum simulators,  either by decreasing circuit-depth overhead in digital approaches or by expanding the interaction sets available in analog hardware. On the digital side, variational quantum eigensolvers~\cite{peruzzo_variational_2014}, which use parametrized circuits to approximate complex ground states, typically focus on static properties and require extensive classical optimization. Pulse-engineering techniques have enabled the construction of higher-order interaction gates involving three or more bodies \cite{Levine2019,kim2022high,Katz2022Nbody,katz2023demonstration,gambetta2020long,wang2001multibit,menke2022demonstration}, which can be integrated into Trotterized sequences. Such higher-order gate constructions extend the digital toolbox but are generally not applicable in analog settings and apply only to limited subsets of qubits. Floquet engineering can expand analog control by varying the strength of native interactions \cite{lee2016floquet,katz2024observing}, but it has limited ability to generate strong higher-order terms beyond the native Hamiltonian \cite{decker2020floquet,choi2020robust,potirniche2017floquet}.

\begin{figure*}[htbp]
\begin{centering}
 \includegraphics[width=6.8in]{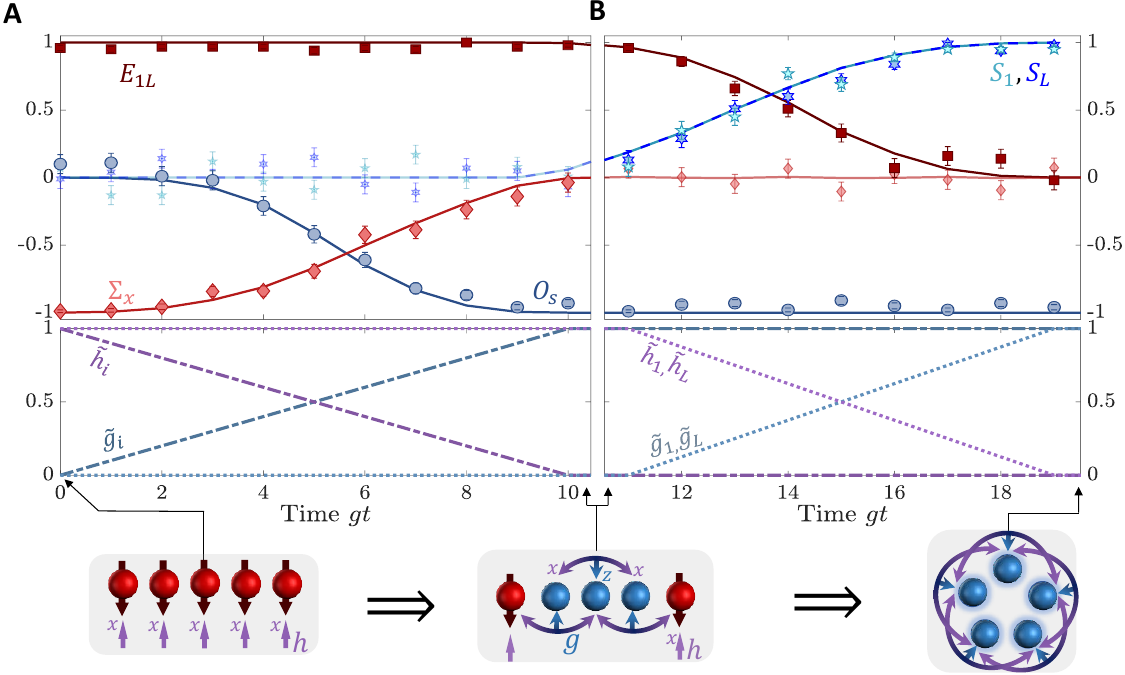}
\par\end{centering}
\centering{}\caption{\textbf{Simulating topological properties of the cluster-field Hamiltonian.} \textbf{a}, Adiabatic preparation of the cluster Hamiltonian ground state with open boundary conditions on a five-spin chain. The bulk fields $h_i$ ($i\neq1,L$) are ramped to zero while the bulk three-body couplings $g_i$ are ramped up; the edge fields $h_1,h_L$ are held constant. As a result, the bulk magnetization $|\Sigma_x|$ decreases, while the nonlocal string order parameter $O_s=\langle\hat{\sigma}_x^{(1)}\hat{\sigma}_y^{(2)}\Pi_{j=3}^{L-2}\hat{\sigma}_z^{(j+2)}\hat{\sigma}_y^{(L-1)}\hat{\sigma}_x^{(L)}\rangle$ between the string edges assumes a non-zero value, heralding the SPT phase. \textbf{b}, Dynamical transition from open to periodic boundary conditions by activating the three-body couplings $g_{1},\,g_{L}$ at the edges while simultaneously deactivating the edge fields $h_{1},\,h_{L}$. The left-right edge correlation $E_{1L}=\langle\hat{\sigma}_x^{(1)}\hat{\sigma}_x^{(L)}\rangle$ decays as edge-localized modes delocalize, while the boundary stabilizers  $ S_1=\braket{\hat{\sigma}_z^{(1)}\hat{\sigma}_x^{(2)}} $ and $S_L=\braket{\hat{\sigma}_x^{(L-1)}\hat{\sigma}_z^{(L)}}$ increase, consistent with uniform couplings around the ring at the final time. Throughout this transition $| O_s|$ stays near unity. The hybrid digital-analog protocol (\ref{fig:Intro}) enables continuous-time control of the effective Hamiltonian, free from Trotter error. Symbols show measured data, while solid lines show theoretical predictions; error bars denote $1\sigma$ binomial uncertainty.\label{fig:Cluster-Field}}
\end{figure*}

Hybrid digital-analog protocols go beyond pulse- and Floquet-engineering by inserting explicit gate operations between analog evolutions. These protocols have been explored as a means to combine the respective advantages of both approaches. Prior implementations have primarily focused on augmenting analog dynamics with single-qubit rotations \cite{arrazola2016digital,parra2020digital,Lei2022,morong2021observation,morong2023engineering,schuckert2025observation} or, more recently, limited two-qubit gates \cite{andersen2025thermalization}. These methods have been applied to improve state preparation \cite{joshi2022,schuckert2025observation}, introduce weak perturbations \cite{Lei2022,morong2021observation}, assist with measurement selectivity \cite{brydges2019probing,huang2020predicting,lu2024digital}, and mitigate errors \cite{morong2023engineering}. While effective for optimizing specific tasks, these methods have not aimed to realize fundamentally new interactions. As a result, the potential of hybrid digital-analog control to construct novel Hamiltonians remains largely underexplored.

Here, we introduce and experimentally demonstrate a hybrid digital-analog protocol that enables quantum simulation of effective Hamiltonians beyond the reach of digital or analog methods alone. Our approach interleaves shallow  layers of entangling gates with segments of analog evolution to  generate coherent, non-native multi-body interactions that are otherwise inaccessible (Fig.~\ref{fig:Hero}). This framework allows us to construct Hamiltonians featuring three- and four-body terms that act simultaneously and non-commutatively with lower-order interactions, implemented without Trotter error. We validate our protocol on a trapped-ion quantum processor by realizing several representative models, including a cluster-field Hamiltonian that contains three-body interaction terms, a cluster-Ising Hamiltonian that realizes a topological spin chain with strong zero modes, and a Hamiltonian with four-body interaction terms. These results expand the class of models accessible to quantum simulators and establish hybrid digital-analog control as a powerful tool for engineering complex dynamics across platforms.

\section*{Results}
\vspace{-5pt}
\subsection*{Hybrid Digital-Analog Quantum Simulation} 
\vspace{-10pt}
We begin by detailing the core mechanism behind our 3-step digital-analog-digital protocol where an analog step is sandwiched between two shallow digital steps that map the native analog Hamiltonian to an effective Hamiltonian with higher-order terms. The protocol consists of three sequential stages: (i) a digital entangling layer that implements a low-depth quantum circuit represented by a unitary operator $\hat D$; (ii) an analog evolution block governed by a native Hamiltonian $\hat H_{\textrm{A}}(t)$, which may be static or time-dependent and can include non-commuting terms; and (iii) a second digital gate layer $\hat D^\dagger$, which reverses the initial encoding in the absence of analog dynamics. This sequence yields an effective time-dependent Hamiltonian $\hat H_{\textrm{eff}}(t)=\hat D^{\dagger}\hat H_{\textrm{A}}(t)\hat D$, enabling the simulation of dynamics governed by multi-qubit and higher-order interactions not present in $\hat H_{\textrm{A}}$. The system evolves \textit{exactly} under the transformed Hamiltonian $\hat H_{\textrm{eff}}$.

To illustrate our protocol, we begin with a simple model of non-interacting spins in a magnetic field, governed by the analog Hamiltonian
$\hat H_{\textrm{A}}(t)=\sum_{j}g_{j}(t)\hat{\sigma}_{z}^{(j)}+\sum_{j}h_j(t)\hat{\sigma}_{x}^{(j)}$.
Here, $\boldsymbol{\hat{\sigma}}^{(j)}$ denotes the vector of Pauli operators for a spin-$\tfrac{1}{2}$ particle $j$.  Conjugation by a shallow entangling circuit $\hat D$ (\ref{fig:Intro}), consisting of up to two entangling gates per spin, yields the effective Hamiltonian \begin{equation}\hat H_{\textrm{eff}}(t) = -\sum_{j=1}^{L}g_{j}(t)\hat{\sigma}_{x}^{(j-1)}\hat{\sigma}_{z}^{(j)}\hat{\sigma}_{x}^{(j+1)}+\sum_{j=1}^{L}h_j(t)\hat{\sigma}_{x}^{(j)},\label{eq:H_cluster_field}\end{equation} which realizes the field-cluster Hamiltonian \cite{wolf2006quantum,smith2022crossing,kalis2012fate}  (with $\hbar=1$). Indices are taken modulo $L$, so that for all integers $k$ and for $n\in\{x,y,z\}$ we have $\hat{\sigma}_{n}^{(k+L)}=\hat{\sigma}_{n}^{(k)}$. Setting $g_1=g_L=g$ retains the boundary three-spin terms that couple site $L$ to site $1$, yielding periodic boundary conditions; setting $g_1=g_L=0$ removes those boundary couplings and produces open boundary conditions for the cluster Hamiltonian. In the zero-field limit ($h_j=0$), this Hamiltonian possesses a $Z_2\times Z_2$ symmetry and hosts a symmetry-protected topological (SPT) phase \cite{verresen2017one}, with applications in quantum information \cite{raussendorf2001one}. Systems governed by field-cluster Hamiltonians exhibit hallmark features of SPT phases including robust edge modes under open boundary conditions and nonlocal string order protected by symmetry. Though analytically tractable, this example illustrates how native one-body terms can be transformed into effective higher-order, nonlocal interactions with minimal gate overhead.

We experimentally realize the effective Hamiltonian of Eq.~(\ref{eq:H_cluster_field}) using a trapped-ion quantum processor based on a linear chain of fifteen $^{171}\mathrm{Yb}^+$ ions. Qubits are encoded in the hyperfine clock states of each ion, and individual optical addressing is achieved using tightly focused Raman beams, enabling full control over both digital gates and analog evolution (see Methods and \ref{fig:Exp_ions_digital_analog} for details). By modulating the local fields within the analog block, we render the effective Hamiltonian explicitly time-dependent.

We first demonstrate our protocol on a five-qubit subsystem by adiabatically preparing a family of ground states of the field-cluster Hamiltonian. The system is initialized in the product state $\ket{\downarrow_x\cdots\downarrow_x}$, which is the ground state of
$\hat H_{\text{eff}}$ in Eq.~(\ref{eq:H_cluster_field}) with $h_{i}=1$ and $g_{i}=0$ for all $i$. After applying the circuit $\hat D$, we deform the effective Hamiltonian by ramping down $h_i$ and ramping up $g_i$ for $2\leq i\leq L-1$ while maintaining the edge fields $g_1=g_L=0$ and $h_{1}=h_{L}=1$ constant. Figure~\ref{fig:Cluster-Field}a shows the measured normalized parameters $\tilde{g}_{i} = {g_i}/{g}$ and $\tilde{h}_{i}={h_i}/{g}$ for the bulk and edge spins as a function of the ramp time $gt$. In the experiment, we compile the ramp into single-qubit gates parameterized by the functions $\tilde{g}_{i}(gt)$ and $\tilde{h}_{i}(gt)$. The ramp gradually suppresses the longitudinal fields in the bulk while activating the three-body interactions consistent with the cluster Hamiltonian under open boundary conditions, culminating in preparation of its ground state at $gt=10$. During this evolution, the bulk magnetization $ \Sigma_x=\sum_{j=2}^{L-1}\langle\hat{\sigma}_x^{(j)}\rangle$ decreases, while the nonlocal string order parameter $ O_\textrm{s}=\langle\hat{\sigma}_x^{(1)}\hat{\sigma}_y^{(2)}\Pi_{j=3}^{L-2}\hat{\sigma}_z^{(j+2)}\hat{\sigma}_y^{(L-1)}\hat{\sigma}_x^{(L)}\rangle$ assumes a nonzero value that is characteristic of the SPT phase \cite{perez2008string,smith2022crossing} (see Methods). 

To probe the controllability of Hamiltonian evolution, we begin with the previously prepared ground state of the cluster Hamiltonian under open boundary conditions (at $gt=10$) and gradually reconfigure the system to impose periodic connectivity. This is achieved by ramping down the longitudinal fields $h_1$ and $h_L$ on the edge spins while simultaneously activating the three-body interaction terms $g_1$ and $g_L$. Here, the digital layer provides a single two-qubit gate between spins 1 and $L$ to complete the ring, effectively stitching the linear spin chain into a closed ring. As this transformation proceeds, the initially edge-localized modes delocalize, evidenced by a decay of the edge-to-edge correlation $E_{1L}=\langle\hat{\sigma}_x^{(1)}\hat{\sigma}_x^{(L)}\rangle$. Concurrently, stabilizer terms  $S_1=\langle\hat{\sigma}_x^{(L)}\hat{\sigma}_z^{(1)}\hat{\sigma}_x^{(2)}\rangle$ and $S_L=\langle\hat{\sigma}_x^{(L-1)}\hat{\sigma}_z^{(L)}\hat{\sigma}_x^{(1)}\rangle$, characteristic of the periodic geometry, emerge. These reflect the formation of uniform three-body couplings across the ring, with complete stitching achieved at the final time, which we chose to satisfy $gt=19$ (Fig.~\ref{fig:Cluster-Field}b). Notably, the nonlocal string-order parameter remains near unity throughout, indicating that the system retains its SPT character during the transition in the geometry of the system. This experiment demonstrates how hybrid digital-analog control can access and manipulate nontrivial Hamiltonians with dynamically programmable geometry.

\begin{figure*}[htbp]
\begin{centering}
 \includegraphics[width=6.1in]{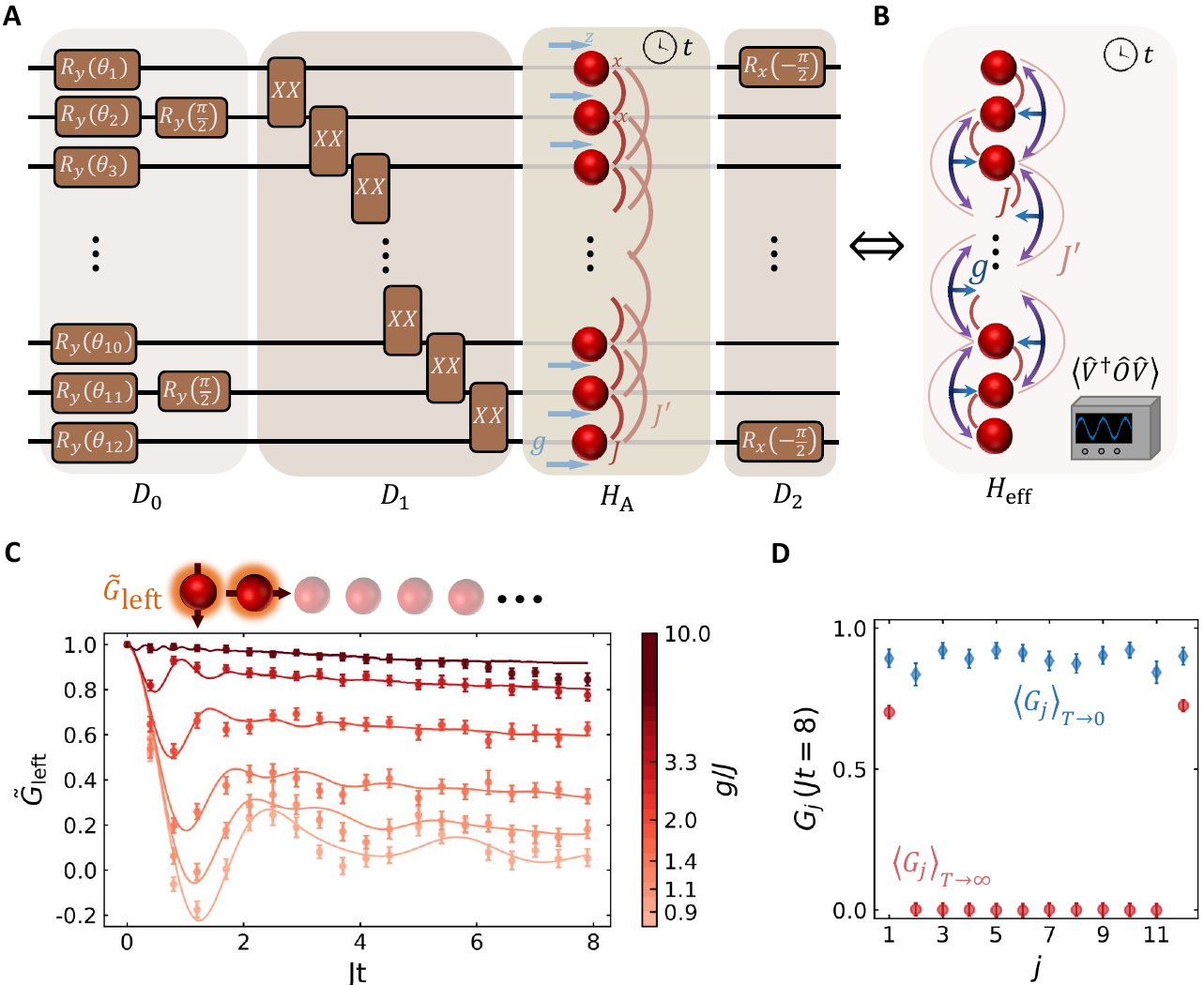}
\par\end{centering}
\centering{}\caption{\textbf{Signatures of strong zero modes.} \textbf{a}, Schematic of the digital-analog protocol used to probe strong zero modes in a cluster-Ising Hamiltonian on a twelve-spin chain. The digital layer $\hat D_0$ prepares product states at varying energies to approximate the infinite-temperature ensemble, followed by a digital layer $\hat D_1$ and analog evolution under the transverse-field Ising Hamiltonian $\hat H_{\textrm{A}}=\sum_{ij}J_{ij}\hat{\sigma}_{x}^{(i)}\hat{\sigma}_{x}^{(j)}+g\sum_{i}\hat{\sigma}_{z}^{(i)}$. A final digital layer $\hat D_2$ completes the sequence, enabling measurement of observables in rotated bases. XX gate between qubits $i$ and $j$ denotes the two-qubit operator $\exp(-i\tfrac{\pi}{4}\hat{\sigma}_{x}^{(i)}\hat{\sigma}_{x}^{(j)})$. \textbf{b}, Effective Hamiltonian dynamics. Choosing $\hat D_2 \neq \hat D_1^\dagger$ implements evolution under the effective Hamiltonian $\hat H_{\textrm{eff}}$ in Eq.~(\ref{eq:H_cluster_Ising}) while enabling efficient measurement of transformed observables $\langle \hat V^\dagger \hat{O} \hat V \rangle$ in place of $\langle \hat{O}\rangle$, where $\hat V = \hat D_2 \hat D_1$. \textbf{c}, Time evolution of the normalized edge correlator $\tilde G_{\textrm{left}}(t)$ at high temperature for varying interaction ratios $g/J$. Solid lines indicate ab-initio numerical predictions that include independently measured qubit dephasing. \textbf{d}, Long-time value of the unnormalized correlator associated with the stabilizer operator at site $j$ for $g/J=3.3$. For the ground state of the cluster Hamiltonian (blue diamonds), the correlator is near unity, consistent with topological protection of the ground state. For the high-temperature state (red circles), it is close to zero except at the edges, consistent with strong zero modes. See Methods for details. In panels \textbf{c}-\textbf{d}, symbols denote measured data, while lines show numerical simulations. Error bars denote the quadrature sum of the $1\sigma$ binomial uncertainties of each state and the $1\sigma$ standard error of the mean over all input states.\label{fig:Cluster-Ising}}
\end{figure*}

\subsection*{Robustness of Strong Zero Modes}\vspace{-10pt}
The cluster Hamiltonian with open boundary conditions supports edge-localized operators that are predicted to retain memory of their initial values over exceptionally long times, even at high energies. This robustness arises from strong zero modes (SZMs): operators that commute with the cluster Hamiltonian and remain approximately conserved under weak perturbations, thereby resisting thermalization over extended timescales \cite{kemp2017,else2017prethermal}. 
Notably, SZMs persist even in highly excited states beyond topological protection, making their signatures potentially observable at very high temperature. Recently, SZMs have also been explored on superconducting-qubit platforms using fully digital protocols that rely on Trotterization~\cite{jin2025observation}.

In a chain of $L$ spins governed by the cluster Hamiltonian with open boundary conditions, four strong zero modes reside at the edges: $\hat{\sigma}_x^{(1)}$, $\hat{\sigma}_x^{(L)}$, $\hat{\sigma}_z^{(1)}\hat{\sigma}_x^{(2)}$, and $\hat{\sigma}_x^{(L-1)}\hat{\sigma}_z^{(L)}$ (see Methods). To probe their stability, we introduce a competing Ising perturbation and study the resulting dynamics under the effective Hamiltonian \begin{equation}\hat H_{\text{eff}} = -\sum_{j=2}^{L-1}g_{j}\hat{\sigma}_{x}^{(j-1)}\hat{\sigma}_{z}^{(j)}\hat{\sigma}_{x}^{(j+1)}+\sum_{i,j=1}^{L}J_{ij}\hat{\sigma}_{x}^{(i)}\hat{\sigma}_{x}^{(j)},\label{eq:H_cluster_Ising}\end{equation} with open boundary conditions, where the couplings $J_{ij}$ decay roughly exponentially with the distance $|i-j|$, and are dominated by nearest-neighbor terms with average strength $J$ (see Methods). Among the four SZMs, two ($\hat{\sigma}_x^{(1)}$ and  $\hat{\sigma}_x^{(L)}$) commute with $H_{\text{eff}}$ and are predicted to remain exact constants of motion. The other two ($\hat{\sigma}_z^{(1)}\hat{\sigma}_x^{(2)}$ and $\hat{\sigma}_x^{(L-1)}\hat{\sigma}_z^{(L)}$) are no longer conserved and become prethermal strong zero modes: they exhibit slow dynamics and persist for long times despite not being exact symmetries (see Methods for the derivation of this fact). 

To probe the dynamical signatures of these prethermal zero modes, we implement a hybrid digital-analog protocol to simulate $\hat H_{\textrm{eff}}$ for $L=12$ spins (Fig.~\ref{fig:Cluster-Ising}\textbf{a},\textbf{b}). A digital layer $\hat D_1$, composed of single- and two-qubit gates, is followed by analog evolution under a transverse-field Ising Hamiltonian $\hat H_{\mathrm{TFIM}}$ with transverse field $g$ and couplings $J_{ij}$. A final digital layer $\hat D_2$ completes the sequence. Under this construction, the dynamics are governed by the effective Hamiltonian $\hat H_{\textrm{eff}}=\hat D_{1}^{\dagger}\hat H_{\mathrm{TFIM}}\hat D_{1}$. The choice of $\hat D_2$ determines the measurement basis: setting $\hat D_2=\hat D_1^\dagger$ yields measurement in the original basis, whereas $\hat D_2\neq \hat D_1^{\dagger}$ enables readout in rotated frames. For an initial state $\ket{\psi_0}$, we define the time-evolved state $\ket{\psi(t)}=\hat{U}_{\textrm{eff}}(t)\ket{\psi_0}$, where $\hat{U}_{\textrm{eff}}(t)$ is the time-evolution operator generated by $\hat H_{\textrm{eff}}$, such that the expectation value of an observable $\hat{O}$ is evaluated as $\langle \hat{O}(t)\rangle=
\bra{\psi(t)} \hat V^\dagger \hat{O} \hat V\ket{\psi(t)}$, where $\hat V = \hat D_2 \hat D_1$. In this work, we focus on the normalized edge correlators $\tilde{G}_{\textrm{left}}(t)\propto\langle\hat{\sigma}_z^{(1)}\hat{\sigma}_x^{(2)}\rangle$ and $\tilde{G}_{\textrm{right}}(t)\propto\langle\hat{\sigma}_x^{(L-1)}\hat{\sigma}_z^{(L)}\rangle$, which probe two prethermal strong zero modes. The proportionality factors are determined by normalizing the initial signal to unity, thereby mitigating state-preparation and measurement errors of about 5\%  (see Methods). 

\begin{figure*}[htbp]
\begin{centering}
 \includegraphics[width=6.8in]{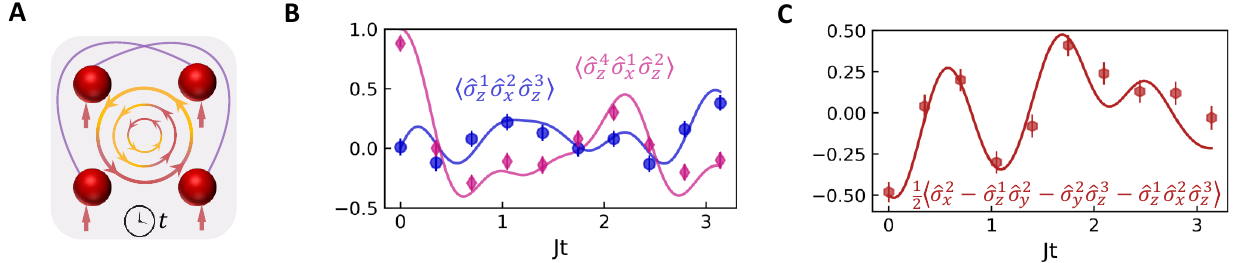}
\par\end{centering}
\centering{}\caption{\textbf{Dynamics under a four-body interaction Hamiltonian.} \textbf{a}, Effective Hamiltonian containing one-body (red arrows), two-body (purple bonds), and four-body (red-yellow rings) terms in a four-spin chain with periodic boundary conditions based on the digital-analog sequence in \ref{fig:Four_body_protocol}. \textbf{b}, Expectation values of individual Pauli strings, each obtained from single-qubit readout. \textbf{c}, Superpositions of multiple Pauli strings from single-qubit readout. The oscillations in \textbf{c} and \textbf{d} show coherent interference between contributions from different Hamiltonian terms (see Methods). Symbols represent measured data; solid lines show numerical simulations without free parameters. Error bars denote $1\sigma$ binomial uncertainties. \label{fig:Four_body}}
\end{figure*}

To probe the high-temperature dynamics of the edge correlators, we prepend to the evolution sequence a randomized single-qubit rotation block $\hat D_0$ that prepares a uniform ensemble of product states $\ket{\psi_0}=\hat{D}_0\ket{0}$ spanning the full energy spectrum, where $\ket{0}$ denotes the fully polarized state with all qubits in $\ket{0}$. Averaging over different inputs approximates infinite-temperature expectation values (see Methods and \cite{schuckert2025observation}). Figure \ref{fig:Cluster-Ising}\textbf{c} shows the measured normalized left-edge correlator as a function of the evolution time $Jt$. This correlator retains a substantial fraction of its initial value even at long times, in close agreement with numerical ab-initio predictions that only include the independently measured qubit dephasing (see Methods). Similar behavior is observed for the right-edge correlator (\ref{fig:Right_edge_correlator}).

To test whether this robustness is unique to the edges at high temperature or persists more generally, we measure the long-time (unnormalized) stabilizer correlators $G_j=\langle \hat{S}_j(t)\rangle\langle \hat{S}_j(0)\rangle$ with $\hat{S}_j=\hat{\sigma}_x^{(j-1)}\hat{\sigma}_z^{(j)}\hat{\sigma}_x^{(j+1)}$ across all sites for both the infinite-temperature ensemble and the cluster-state ground state. Figure~\ref{fig:Cluster-Ising}\textbf{d} shows that, for the ground state (blue diamonds), $G_j$ remains near unity across the chain. For the infinite-temperature case (red circles), $G_j$ remains near zero in the bulk and finite only at the edges, consistent with strong zero modes. Full time traces for all sites are shown in \ref{fig:gnd_state_correlator} and \ref{fig:infintie_temp_state_correlator}. Together, these observations highlight the topological protection of the cluster ground state and the strong zero-mode character of the edge operators at infinite temperature.

\subsection*{High-Order Interaction Hamiltonians}\vspace{-10pt}
The hybrid digital-analog method extends beyond the cluster-Hamiltonian construction to realize multi-qubit interactions of arbitrary order. Each two-qubit entangling gate in the digital layer $\hat D$ acting on a given qubit can increase the order of the corresponding effective Hamiltonian term by one, enabling systematic construction of $k$-body couplings from native interactions via gate application (see Methods). These engineered terms appear with non-perturbative strength and can act simultaneously with other interactions in $\hat H_{\mathrm{A}}$. Even at minimal circuit depth, this approach can yield effective Hamiltonians containing non-commuting three-, four-, or higher-body interactions, substantially broadening the accessible model space.

In the examples above [Eqs.~(\ref{eq:H_cluster_field})-(\ref{eq:H_cluster_Ising})], two entangling gates per qubit transform one-body fields into effective three-body interactions. To illustrate the versatility of the method and its ability to generate higher-order terms, we construct an effective Hamiltonian $\hat H_{\text{eff}}=\hat D_1^\dagger \hat H_{\mathrm{A}}\hat D_1$ given by\begin{equation}
\begin{aligned}
     \hat H_{\text{eff}} =& J\sum_{j=1}^{L-1}\hat{\sigma}_{x}^{(j-1)}\hat{\sigma}_{y}^{(j)}\hat{\sigma}_{y}^{(j+1)}\hat{\sigma}_{x}^{(j+2)}\\
     +& J'\sum_{j=1}^{L-2}\hat{\sigma}_{z}^{(j)}\hat{\sigma}_{z}^{(j+2)}+h\sum_{j=1}^{L}\hat{\sigma}_{x}^{(j)},
\end{aligned}
\label{eq:H_four_body}
\end{equation} which combines non-commuting one-, two-, and four-body terms as illustrated in Fig.~\ref{fig:Four_body}\textbf{a}. Here, the operators are defined with periodic boundary conditions, $\hat{\sigma}_{n}^{(k+L)}=\hat{\sigma}_{n}^{(k)}$ for $k\geq0$.

The four-body Hamiltonian term in the first line is for example related to the Wen plaquette model~\cite{Wen2003}---a foundational model of $Z_2$ topological order and important in topological quantum error correction---and via a Jordan-Wigner transformation to four-body Majorana fermion interactions~\cite{Affleck2017}.

We experimentally implement this Hamiltonian in the minimal configuration of a four-spin chain with periodic boundary conditions, which is the minimal example capturing the rich dynamics induced by the four-body terms. The protocol begins with the preparation circuit $\hat D_0$, followed by the entangling layer $\hat D_1$ and evolution under the native TFIM Hamiltonian with open boundary condition for the analog block (\ref{fig:Four_body_protocol}) with nearest-neighbor terms with strength $J$ and next-nearest-neighbor terms with strength $J'$ and negligible contribution of higher order terms. The dynamics are probed in the effective Hamiltonian frame by selecting different analysis circuits $\hat D_2$, enabling efficient access to a range of observables, including high-order Pauli operators (Fig.~\ref{fig:Four_body}\textbf{b}) and linear combinations of multiple Pauli strings (Fig.~\ref{fig:Four_body}\textbf{c}) (see Methods). The measured results agree with theoretical modeling. This example illustrates the flexibility of the hybrid scheme in realizing nontrivial, non-commuting interactions beyond the two-body limit using shallow circuits in combination with native analog resources.

\section*{Discussion}
This work demonstrates that hybrid digital-analog control can realize effective Hamiltonians featuring higher-order and non-commuting interactions. It achieves this without the circuit-depth overhead of digital Trotterization and enables access to interaction terms that are not natively available in analog settings. These results establish digital-analog control as a practical framework for exploring complex quantum dynamics beyond the reach of either paradigm alone. 

The effective analog evolution retains key features of the native analog block: it preserves the non-commutativity of the underlying Hamiltonian terms and shares its eigenvalue spectrum. In the digital-analog protocol, the digital unitaries $\hat D_1$ and $\hat D_2$ applied before and after the analog evolution define an effective evolution of the transformed state $\hat D_1\ket{\psi_0}$ under $\hat H_{\textrm{A}}$, followed by measurement in the computational basis rotated by $\hat D_2^{\dagger}$. Equivalently, this procedure simulates the evolution of $\ket{\psi_0}$ under the effective Hamiltonian $\hat H_{\textrm{eff}} = \hat D_1^\dagger \hat H_{\textrm{A}} \hat D_1$, with final measurement in the computational basis rotated by $\hat{V}^{\dagger}$. The choice $\hat D_2 \neq \hat D_1^{\dagger}$ provides additional flexibility to tailor the measurement basis beyond the computational frame.

While $\hat H_{\textrm{eff}}$ and $\hat H_{\textrm{A}}$ are related by a basis transformation, attempting to reproduce arbitrary dynamics of generic observables under $\hat H_{\textrm{eff}}$ using only analog evolution under $\hat H_{\textrm{A}}$ is, in general, extremely challenging. Although the circuits $\hat D_1$ and $\hat D_2$ are shallow, their introduction can be crucial when the analog time-evolution operator $\hat U_{\textrm{A}}(t)=e^{-i\hat H_{\textrm{A}}t}$ induces significant operator spreading. A state such as $\hat U_{\textrm{A}}(t)\hat D_1\ket{\psi_0}$, while simple to prepare at $t=0$, can evolve into a highly entangled state that cannot be approximated by product states or, in general, be efficiently sampled in any fixed local measurement basis. In fact, if $\hat D_1$ is a constant-depth 2D IQP (instantaneous quantum polynomial) circuit (which we assume not to be implementable by analog-only evolution), it is classically hard to sample even at $t=0$~\cite{Bremner2016}. Likewise, a weight-$w$ observable $\hat{O}$ consisting of a single Pauli string maps to in general an exponential-in-$w$ number of Pauli strings in the analog frame, $\hat{\tilde{O}}=\hat D_2^{\dagger}\hat{O}\hat D_2$. Since we assume that the analog simulator can only apply single-qubit gates to change the measurement basis, each of these Pauli strings at least naively requires a different experiment. In this case, estimating $\langle \hat{\tilde{O}}(t) \rangle$ from analog-only evolution  may require exponentially many measurement settings. In contrast, the digital-analog protocol applies $\hat D_2$ physically, enabling the measurement of $\hat O$ with such a single measurement setting. Therefore, our technique extends the reach of analog quantum simulators to effective Hamiltonians and observables that are otherwise inaccessible or extremely challenging in purely analog settings. In this work, it enables, for example, efficient access to strong zero-mode dynamics in the cluster-Ising Hamiltonian, including regimes far from the ground state. 

Implementing the protocol requires the ability to rapidly switch between qubit-selective digital gates and global analog interactions within a single experimental sequence. In trapped-ion systems, this capability relies on parallel frequency and amplitude control of Raman beams at each ion site, which sets the coupling to the motional modes that mediate interactions between spins/qubits and allows programmable pulse shaping. In our setup, this is achieved through a programmable array of individually addressed beams operating simultaneously across the full ion chain. A further requirement is maintaining phase coherence between digital and analog segments. We use the same physical laser fields for both modes to minimize uncontrolled phase shifts, applying phase corrections to compensate for local oscillator frequency changes between segments (see Methods). 

The digital-analog protocol can be included as a modular building block within larger quantum circuits, or adapted to a wide range of physical platforms that support different native analog Hamiltonians. Examples include spin-boson or bosonic simulators in trapped-ion systems~\cite{whitlow_quantum_2023,katz_programmable_2023,so_quantum_2025}, as well as systems of Rydberg atom arrays~\cite{labuhn_tunable_2016,bernien_probing_2017,cao_multi-qubit_2024} and superconducting qubits~\cite{wang_experimental_2022,crane2024}. The core ingredients, namely native analog evolution combined with programmable shallow entangling circuits, make the approach widely applicable across quantum computing architectures.

\clearpage

\part*{\centerline{Materials and Methods}}

\renewcommand{\figurename}{}
\newcounter{extended_data_fig}
\setcounter{extended_data_fig}{1}
\renewcommand{\thefigure}{Extended Data Fig.~\arabic{extended_data_fig}}

\section*{Experimental Details}
\vspace{-7pt}
In this Section, we describe the experimental setup and settings used to implement the hybrid digital-analog protocol on a trapped-ion quantum processor. Experiments are conducted on a universal trapped-ion quantum computer consisting of $15$ $^{171}$Yb$^{+}$ ions confined in a room-temperature Sandia HOA-2.1.1 surface linear Paul trap \cite{maunz_high_2016}. Trapping electrodes are driven with amplitude of approximately 200 V at $36.064$ MHz, resulting in secular radial ion motion frequency of approximately $3.19$ MHz. The qubit states are encoded in the $2S_{1/2}$ ground hyperfine ``clock'' states, $|0\rangle = |F=1;m_{F}=0\rangle$ and $|1\rangle = |F=0;m_{F}=0\rangle$, with a first-order magnetic-insensitive splitting of $\omega_{0}=2\pi\times 12.642818614(10)$ GHz at the operating field of 0.457 mT. All qubits are initialized to $|1\rangle$ via optical pumping, and motional states of lower radial modes are prepared near the ground state via Raman sideband cooling. Coherent qubit control is achieved via a global-addressing $355$ nm beam together with $13$ individual-addressing $355$ nm beams that drive Raman transitions between the qubit states. The ions are nearly equally-spaced by  about $3.8$ $\mu$m, ensuring alignment with the centers of the individual-addressing beams and multimode-fiber coupled photomultiplier tubes (PMTs) that are used for state readout.

Amplitude- and phase- control of the addressing Raman beams is achieved via acousto-optic modulators (AOMs) from the L3 Harris Corporation. Radio-frequency (rf) control waveforms are produced by rf system-on-a-chip (RFSoC) evaluation boards (ZCU111 from Xilinx Corporation), driven by the Octet control firmware  \cite{lobser2023jaqalpaw}. The Octet firmware allows independent phase-tracking of all the output waveforms.

The hybrid quantum evolution consists of digital single-qubit 
and two-qubit 
gates, and analog Hamiltonian evolution. Single-qubit gates are performed using Gaussian-shaped composite SK1 pulses to minimize pulse amplitude errors \cite{brown_arbitrarily_2004, miao_probing_2025}. Two-qubit gates and Hamiltonian evolution are implemented through M\o lmer-S\o rensen (MS) interaction that couples spins to the lower-frequency set of radial modes. The required bichromatic drive is achieved by applying two tones to the global Raman beam.

Two-qubit gates are driven by a MS interaction operated in the near-resonant regime, with the bichromatic detuning optimized to lie near the midpoint between pairs of adjacent modes. The gate pulse is configured as a segmented amplitude-modulated waveform with total duration set to minimize residual spin-motion entanglement \cite{egan_fault-tolerant_2021}. Varying gate angles are realized by tuning the individual beam intensity. The gates are executed sequentially to bypass undesired couplings. We calibrate the individual beam amplitude and the imbalance between the red and blue tones of the global beam and account for the residual light shift during the interaction as described in Ref.~\cite{miao_probing_2025}.

Analog Hamiltonian evolution is implemented with the MS interaction in the dispersive regime, in which the bichromatic detuning is set to be far-detuned from the relevant motional modes. We adjust the individual beam intensities so that the experimentally measured nearest-neighbor spin-spin coupling is leveled to desired values and the second-nearest-neighbor coupling is uniform \cite{Lei2022}. Each ion pair is calibrated individually, with the two participating spins initialized in opposite orientations to suppress the detuning arising from the common-mode light shifts. We measure the light shift by scanning the frequency of each individual beam during the interval between the two pulses that rotate the spins in and out of the interaction basis \cite{schuckert2025observation}.

Motion of the ion chain in its lowest-frequency axial mode causes unwanted modulation of the amplitude of the individual-addressing laser beams as seen by the ions \cite{huang_comparing_2024}. 
In the implementation of the analog Hamiltonian on a long chain, we suppress the coupling to the first axial sideband of the radial modes by selecting a motional detuning that is centered between two adjacent sidebands of radial modes. 

In the measurement of nontrivial zero-modes, to counteract the dephasing, which is primarily caused by uncompensated light shifts, we apply an asymmetric Carr-Purcell-Meiboom-Gill (CPMG) echo sequence during the analog evolution \cite{morong2023engineering}. We divide the evolution into two equal-duration segments, apply $\hat R_{x}(\pi)$ sequentially to all involved qubits between the segments, reverse the transverse field direction for the second block, and finally apply $\hat R_{x}(\pi)$ to all qubits. In the analog evolution, the relative Rabi frequencies of the 12 individual beams are set as [0.904, 0.605, 0.515, 0.466, 0.499, 0.524, 0.534, 0.512, 0.504, 0.456, 0.508, 0.588], leveling the nearest-neighbor coupling strength and yielding an average coupling strength $J\approx 2\pi \times 200$ Hz. The motional detuning is set $\mu=-2\pi\times 60$ kHz relative to the lowest radial mode.

For simulating the evolution under high-order interaction Hamiltonian, we set $\mu=-2\pi\times 51$ kHz with calibrated relative Rabi frequencies of the 12 indiviual beams as [0.844, 0.580, 0.444, 0.438, 0.452, 0.485, 0.476, 0.497, 0.440, 0.437, 0.486, 0.523]. The leveled average nearest-neighbor coupling strength is $J\approx 2\pi \times 200$ Hz, and the average second-nearest-neighbor coupling strength $\tilde{J}\approx 2\pi \times 78$ Hz. We do not apply the echo sequence during this evolution.  

\section*{Prethermal strong zero modes in the cluster-Ising model}
In this Section, we outline the theoretical framework for strong zero modes in the cluster-Ising Hamiltonian, following mainly Ref.~\cite{kemp2017}.

\subsection*{Strong zero modes in the cluster Hamiltonian}
We first discuss the zero modes of the cluster Hamiltonian. 

We define it as
\begin{equation}
	 \hat H_\mathrm{cluster} = -g\sum_{i=2}^{L-1} \hat{\sigma}_x^{(i-1)} \hat{\sigma}_z^{(i)} \hat{\sigma}_x^{(i+1)}.
\end{equation}
This Hamiltonian exhibits sublattice parity symmetries\begin{align}
	\hat {\mathcal{F}}_1 &= \hat{\sigma}_z^{(1)} \hat{\sigma}_z^{(3)}\cdots \hat{\sigma}_z^{(L-1)}, \\
	\hat {\mathcal{F}}_2 &= \hat{\sigma}_z^{(2)}\hat{\sigma}_z^{(4)}\cdots \hat{\sigma}_z^{(L)}.
\end{align}
Here, we consider an even system size $L$. For open boundary conditions, the cluster model exhibits four edge modes,
\begin{equation}
	\hat{\sigma}_x^{(1)},\; \hat{\sigma}_x^{(L)},\; \hat{\sigma}_z^{(1)} \hat{\sigma}_x^{(2)},\; \hat{\sigma}_x^{(L-1)}\hat{\sigma}_z^{(L)}.
\end{equation}
All of these edge modes are in fact \emph{strong} zero modes, i.e. their infinite-temperature autocorrelator \begin{equation}
	G(t)=\frac{1}{2^{L}} \mathrm{Tr} \left(\hat A(t) \hat A(0) \right)
\end{equation}
is unity for all times. The name ``zero mode'' originates from nanowires, in which Majorana edge modes would appear as a peak in the STM (scanning tunneling microscope) spectrum at zero frequency, which corresponds to the Fourier transform of $G(t)$. This zero frequency peak therefore corresponds to a non-zero signal as $t\rightarrow\infty$. ``Strong'' zero mode means that this feature is even visible at infinite temperature, i.e. not just in the ground state.

The ``strongness'' of the edge mode operators is connected to the following three properties~\cite{kemp2017}:\\
(\textit{i}) They anticommute with either $\hat {\mathcal{F}}_1$ or $\hat {\mathcal{F}}_2$.\\
(\textit{ii}) They square to identity, i.e.~they are normalized.\\
(\textit{iii}) They commute with the Hamiltonian.

From these three properties one can easily show that acting with one of the zero mode operators on an eigenstate of the Hamiltonian within one $\hat {\mathcal{F}}_1,\,   \hat {\mathcal{F}}_2$ symmetry sector will yield an eigenstate in another sector with the \emph{same} eigenenergy. From this fact one can deduce that $G(t)=1$ for all times $t$ absent perturbations to the cluster Hamiltonian~\cite{kemp2017}.

\subsection*{Strong zero modes in the presence of nearest-neighbor Ising interactions}

We now discuss the stability of the zero modes in the cluster Hamiltonian when adding nearest-neighbor Ising interactions
\begin{equation}
	\hat H_\mathrm{Ising} = - J\sum_{i=1}^{L-1} \hat{\sigma}_x^{(i)} \hat{\sigma}_x^{(i+1)}
\end{equation}
to the Hamiltonian. Their presence mean that the sublattice parity symmetries are replaced by $\hat {\mathcal{F}}=\hat {\mathcal{F}}_1\hat {\mathcal{F}}_2$. 
The strong zero modes $\hat{\sigma}_x^{(L)},\,\hat{\sigma}_x^{(1)}$ remain unchanged as they still commute with the Hamiltonian. This is not the case for $\hat{\sigma}_z^{(1)} \hat{\sigma}_x^{(2)}$ and $ \hat{\sigma}_x^{(L-1)}\hat{\sigma}_z^{(L)}$. However, for both $\hat{\sigma}_z^{(1)} \hat{\sigma}_x^{(2)}$ and $ \hat{\sigma}_x^{(L-1)}\hat{\sigma}_z^{(L)}$, we can still construct a new operator $\hat \Psi$ which commutes with the Hamiltonian and which is localized around the edges with an iterative procedure shown in Ref.~\cite{kemp2017} for a related model.

We start the process by defining $\hat \Psi^{(0)}=\hat{\sigma}_z^{(1)} \hat{\sigma}_x^{(2)}$. The commutator of this operator with the Ising Hamiltonian is
\begin{equation}
	\left[\hat \Psi^{(0)},\hat H_\mathrm{Ising} \right] = -2iJ\hat{\sigma}_y^{(1)}.
\end{equation}
In order to compensate for this term, we add $\hat \Psi^{(1)}=\frac{J}{g}\hat{\sigma}_z^{(1)} \hat{\sigma}_z^{(2)}\hat{\sigma}_x^{(3)}$ to the zero mode operator as its commutator with $\hat H_\mathrm{cluster}$ cancels the above term. However, $\hat{\Psi}^{(1)}$ again does not commute with $\hat H_\mathrm{Ising}$. This non-commuting term can be compensated by adding a term $\hat \Psi^{(2)}=\left(\frac{J}{g}\right)^2\hat{\sigma}_z^{(1)} \hat{\sigma}_z^{(2)}\hat{\sigma}_z^{(3)}\hat{\sigma}_x^{(4)}$. Iterating this procedure, we find the new operator
\begin{equation}
	\hat \Psi= N \sum_{j=0}^{L-2} \left( \frac{J}{g}\right)^j \hat{\sigma}_x^{(j+2)} \Pi_{k=1}^{j+1} \hat{\sigma}_z^{(k)}.
\end{equation}
In order to enforce property (ii) of the strong zero mode conditions, the normalization is set to 
\begin{equation}
	N^2=\frac{1-\left( \frac{J}{g}\right)^2}{1-\left( \frac{J}{g}\right)^{2L-3}}.
\end{equation}
This new operator $\hat \Psi$ commutes with the Hamiltonian up to an error $[\hat \Psi,\hat H]=N J (J/g)^{L-2} \hat{\sigma}_x^{(L-1)} \Pi_k^{L-1} \hat{\sigma}_z^{(k)}$, which is exponentially small in the system size for $J<g$. Therefore, and because it also anticommutes with $\hat{\mathcal{F}}$, $\hat{\Psi}$ is a strong zero mode in the infinite-system-size limit as long as $J/g <1$.

The presence of this new strong zero mode means that, even in the presence of nearest-neighbor Ising interactions, the infinite-temperature autocorrelator for the operator $\hat A=\hat{\sigma}_z^{(1)}\hat{\sigma}_x^{(2)}$ does not decay to zero and saturates at $N^2$~\cite{kemp2017}.

\subsection*{Prethermal strong zero modes with next-nearest-neighbor interactions}

\begin{figure}
	\includegraphics{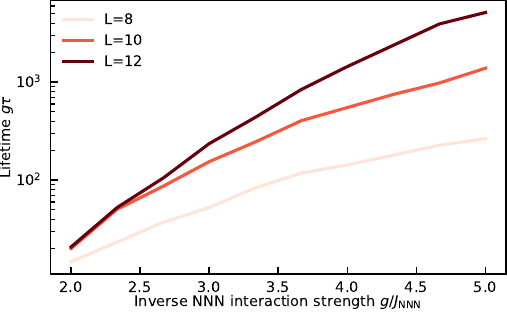}
	\caption{Lifetime $\tau$ of the operator $\hat A=\hat{\sigma}_z^{(1)} \hat{\sigma}_x^{(2)}$ defined by $G(t=\tau)=0.3$ as a function of next-nearest-neighbor interaction strength $J_\mathrm{NNN}/g$. Nearest-neighbor coupling is fixed to $J/g=0.3$.\label{fig:prethermal}}
\end{figure}

We now discuss the influence of beyond-nearest-neighbor interactions on the zero modes. When arbitrary interactions are added to the Hamiltonian, the lifetime of the strong zero mode, measured by the decay rate of $G(t)$, is in general finite. However, 
in Ref.~\cite{else2017prethermal}, it was shown that this lifetime can be \emph{exponentially large} in the perturbation strength due to the phenomenon of prethermalization. As the assumptions necessary for this argument to work are in our case fulfilled, in particular that the interactions are exponentially decaying, we expect this argument to hold. In \ref{fig:prethermal}, we give numerical evidence that this is the case. We add a term $\hat H_\mathrm{NNN}= - J_\mathrm{NNN}\sum_{i=1}^{L-2} \hat{\sigma}_x^{(i)} \hat{\sigma}_x^{(i+2)}$ to the Hamiltonian and find that the autocorrelator of the edge mode $\hat{\sigma}_z^{(1)} \hat{\sigma}_x^{(2)}$ indeed shows an exponentially large lifetime in the next-nearest-neighbor (NNN) interaction strength $J_\mathrm{NNN}$.

\subsection*{Sampling error of infinite-temperature observables}\vspace{-7pt}
We discuss here our scheme to evaluate the infinite-temperature correlator by product-state sampling. 

The scheme is based on choosing a product-state basis $\ket{a}$ such that $\hat A(0)\ket{a}=a \ket{a}$. Using this choice, we can
evaluate $G(t)$ as
\begin{align}
    G(t) &= \frac{1}{2^L} \mathrm{Tr}\left(\hat A(t) \hat A(0) \right)\\
    &= \frac{1}{2^L} \sum_a \braket{a|\hat A(t) \hat A(0)|a}\\
    &= \frac{1}{2^L} \sum_a a \braket{a|\hat A(t)|a}.
\end{align}
We also choose the $\ket{a}$ to be an eigenstate of one of the Pauli strings $\hat P$, i.e. a single tensor product of the Pauli operators $\hat\sigma^x, \hat\sigma^y, \hat\sigma^z$ or the identity.
At first sight, this means that we need to now measure $\braket{a|\hat A(t)|a}$ for all $\ket{a}$, which would be prohibitive as there are $2^L$ such states. Instead, as already done previously~\cite{joshi2022,schuckert2025observation}, we sample the $\ket{a}$ uniformly randomly, i.e.~we draw $S$ samples and each sample is uniformly randomly sampled from our $2^L$ states. We define 
\begin{equation}
    G_S(t)=\frac{1}{S} \sum_{s=1}^S a_s \braket{a_s|\hat A(t)|a_s}.
\end{equation}
Note that we assume here that we take infinitely many shots per state as, for finitely many shots, additional quantum projection noise appears.
In the limit of a large number of samples, it is clear that this scheme produces the correct result because it is an unbiased estimator: because of the uniform sampling, every state is as likely as the other, and $\lim_{S\rightarrow\infty} G_S(t)=G(t)$. More interesting is the question of how quickly the sampling converges. To answer this, it is useful to consider $X_s=\braket{a_s|\hat A(t)\hat A(0)|a_s}$ as a random variable (sampled by sampling different $\ket{a_s}$) over which we average. The variance $\mathrm{Var}(G_S)=\mathrm{Var}\left(\frac{1}{S}\sum_s X_s\right)$ is then given by
\begin{equation}
    \mathrm{Var}(G_S)=\frac{\sigma^2}{S}
\end{equation}
due to the linearity of the variance, where $\sigma^2=\mathrm{Var}(X_s)$. As usual, the root mean squared error therefore decreases as $1/\sqrt{S}$ with the number of samples. 

We next consider how $\sigma^2$ depends on the system parameters. To answer this, note that the variance can be expressed as
\begin{align}
    \sigma^2&=\frac{1}{2^L}\mathrm{Tr}\left(\hat \Delta^2\right) -G(t)^2 \\&\leq \frac{1}{2^L}\mathrm{Tr}\left(\hat \Delta^2\right),
\end{align}
with $\hat \Delta=\sum_{a=1}^{2^L} \ket{a} \bra{a}\hat A(t)\hat A\ket{a}\bra{a}$. To evaluate this expression, we expand $\hat A(t)\hat A$ in the Pauli basis,
\begin{equation}
    \hat A(t)\hat A(0)=\sum_P c_P(t) \hat P,
\end{equation}
where $\hat P$ are all the $4^L$ Pauli strings and $c_P=\mathrm{Tr}\left(\hat P\hat A(t)\hat A(0)\right)$. Because $\hat \Delta$ is diagonal in the $\ket{a}$ basis, we find that
\begin{equation}
    \hat \Delta = \sum_{P\,\mathrm{diag.}}c_P(t) \hat P.
\end{equation}
Note that the sum is only over ``diagonal'' $\hat P$, i.e.~those $\hat P$ with $\braket{a|\hat P|a}=1$ for one of the states $\ket{a}$.

Because of the orthogonality of Pauli strings, this furthermore implies that
\begin{equation}
    \frac{1}{2^L}\mathrm{Tr}\left(\hat \Delta^2\right)=\sum_{P\,\mathrm{diag.}}c_P^2(t).
\end{equation}

It is expected that, for chaotic, scrambling dynamics, an initially local operator $\hat A$, such as the examples we measured in our experiments, spreads uniformly over all non-trivial Pauli strings within the Lieb-Robinson lightcone, e.g.~in 1D a region $L(t) \leq c_\mathrm{LR} t$, where $c_\mathrm{LR}$ is the Lieb-Robinson velocity. There are $M=4^{L(t)}-1$ such strings, such that $c_P^2\propto 1/M$. However, there are only $m=2^{L(t)}-1$ ``diagonal'' strings which contribute to $\sigma^2$. For example, if $a$ was the z-basis, only identity and $\hat{\sigma}^z$ operators are part of the diagonal strings. Therefore,
\begin{align}
    \sum_{P\,\mathrm{diag.}}c_P^2&\approx \frac{m}{M} \\&= 2^{-L(t)}\frac{1-2^{-L(t)}}{1-4^{-L(t)}}\\&\approx 2^{-L(t)}.
\end{align}
Therefore, there are two time regimes of the variance: initially, while decreasing exponentially with time, it does not decrease with system size. After the lightcone hits the system boundaries, the variance is approximately time-independent and decreases exponentially with system size. This means that, for early times, more samples may be required the product state sampling to converge. This behavior needs to be contrasted with the behavior when sampling Haar-random states: in that case, the variance decreases exponentially with system size for all times. Moreover, conservation laws may render the operator spreading slower than linear, further slowing the averaging down~\cite{Keyserlingk2018}. In addition to the discussion here, it is important to note that the variance $\sigma^2$ does not increase with system size for operators with support on a single site simply by using the boundedness of the Pauli operators~\cite{seetharam2023}.

\subsection*{Sampling advantage of digital-analog}

While the general sampling advantage of operators was discussed in the main text, in this section we discuss whether our digital-analog scheme can be advantageous in estimating the infinite-temperature edge correlator $G_{\textrm{left}}(t)$. 

For this observable, one is free to choose the basis to sample from in the bulk, because $\hat A(0)$ in our case only acts on the edge. If we were to sample x product states in the bulk, then the gates in the middle of the system would not have to be applied, and the protocol could be done fully in analog mode. Therefore, a comparison of sampling in the x basis or z basis is also a comparison of a pure-analog scheme and a digital-analog scheme because the z basis requires application of two-qubit gates whereas the x basis scheme does not. 

\begin{figure}
    \centering
    \includegraphics{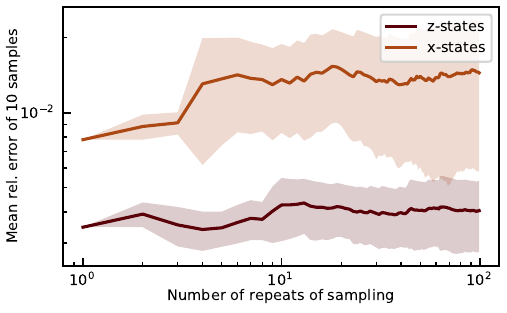}
    \caption{\textbf{Comparison of different sampling bases for determining $G(t)$ for $\hat A=\hat \sigma_z^{(1)}\hat \sigma_x^{(2)}$.} We plot the numerically-obtained relative error of $10$ samples for the edge autocorrelator for $g/J=0.5$, averaged over times $Jt=[0,10]$ for $L=12$. The x-axis denotes the number of repeats of this sampling, with the solid lines indicating the mean of the estimate and the bands its standard deviation.}
    \label{fig:sampling}
\end{figure}

We numerically find that the sampling variance of $G(t)$ is almost an order of magnitude smaller when sampling z product states in the bulk, see \ref{fig:sampling}, which requires gates and therefore our digital-analog scheme. This indicates that the prefactor in the scaling of $\sigma^2$ with the number of samples is smaller for the z basis than the x basis, possibly relating to more strings being diagonal in the operator expansion of $\hat A(t)\hat A$.

Additionally, brickwork circuits of depth $\mathcal{O}(N)$~\cite{Brand_o_2016} can prepare unitary 2-designs, for which the variance for sampling our observable decreases exponentially with system size for all times. While this motivates a possible application of our digital-analog technique---preparing states from a 2-design digitally and then time-evolving using analog time evolution---we note that 2-designs can also be obtained by purely analog time evolution at the cost of using ancilla qubits~\cite{PRXQuantum.4.010311}.

\subsection*{Experimentally measured correlators}
In experiment, we consider the correlators  \begin{equation}G_j(t)=\langle \hat{S}_j(t)\rangle\langle \hat{S}_j(0)\rangle\label{eq:correlator_Gj}\end{equation} with the average taken over all sampled states. The stabilizer operator is $
\hat{S}_j=\hat{\sigma}_x^{(j-1)}\hat{\sigma}_z^{(j)}\hat{\sigma}_x^{(j+1)}$  in the bulk ($2\leq j\leq 11$) and $\hat{S}_1=\hat{\sigma}_z^{(1)}\hat{\sigma}_x^{(2)} $, $\hat{S}_L=\hat{\sigma}_x^{(L-1)}\hat{\sigma}_z^{(L)}$ at the edges (associated with an open boundary chain spin considered in Fig.~\ref{fig:Cluster-Ising}. Note that for periodic boundaries the stabilizer at the edge take instead the same three-body structure as in the bulk). We identify $G_1\equiv G_{\textrm{left}}$ and $G_L\equiv G_{\textrm{right}}$. We define the normalized version of the edge correlators  $\tilde{G}_{\text{left}}(t) = {G(t)}/{G_{\text{left}}(0)}$ and $\tilde{G}_{\text{right}}(t) = {G(t)}/{G_{\text{right}}(0)}$ with the experimentally measured values $G_{\text{left}}(0)=0.908(10)$ and $G_{\text{right}}(0)=0.897(8)$. This normalization partially accounts for effects associated with imperfect state preparation and measurement errors of the qubits.

\section*{High-order interactions}
In this Section, we describe how high-order interactions terms are engineered using the hybrid digital-analog protocol.

Our hybrid digital-analog protocol enables nonperturbative construction of higher-order interaction terms within the effective Hamiltonian $\hat H_\mathrm{eff}$. To illustrate the mechanism, consider a situation in which the analog block with Hamiltonian $\hat H_\mathrm{A}$ contains a term of the form $\hat \alpha \hat{\sigma}_z^{(i)}$, where $\hat \alpha$ is a general operator that may involve spin operators on other sites but commutes with $\hat{\sigma}_x^{(j)}$. We apply a shallow digital circuit containing the two-qubit gate $\hat T_{ij} = \exp(-i\theta \hat{\sigma}_x^{(i)} \hat{\sigma}_x^{(j)})$, and examine the conjugated operator $\hat T_{ij}^\dagger \hat \alpha \hat{\sigma}_y^{(i)} \hat T_{ij}$, associated with the effective Hamiltonian evolution.

Using operator identities and the fact that $(\hat{\sigma}_x^{(j)})^2 = \mathbb{1}$ for spin $\tfrac{1}{2}$, we obtain the exact transformation:
\begin{equation}
\hat T_{ij}^\dagger \hat{\alpha} \hat{\sigma}_z^{(i)} \hat T_{ij}
= \cos(2\theta)\, \hat \alpha \hat{\sigma}_z^{(i)} + \sin(2\theta)\, \hat \alpha \hat{\sigma}_y^{(i)} \hat{\sigma}_x^{(j)} \,.
\end{equation}
This shows that a two-body term in $\hat H_\mathrm{A}$, when conjugated by the entangling gate $\hat T_{ij}$, can generate a new term of higher interaction order involving an additional qubit. In the case where $\hat \alpha$ does not act on qubit $j$, choosing a fully entangling gate with $\theta = \pi/4$ eliminates the original term and yields a higher-order interaction operator $\hat \alpha \hat{\sigma}_y^{(i)} \hat{\sigma}_x^{(j)}$ instead. This method enables direct synthesis of non-perturbative interaction terms.

\section*{Circuit Compilation}
In this section, we detail here the compiled circuits used for the experiments in the main text. 

For the cluster-field Hamiltonian experiment (Fig.~\ref{fig:Cluster-Field}), the preparation block was \begin{equation}
\hat D_1 = \left(\prod_{j=1}^{5} e^{-i\tfrac{\pi}{4} \hat{\sigma}_x^{(j)}\hat{\sigma}_x^{(j+1)}}\right)\prod_{j=1}^{5}\hat {\textrm{H}}_j,
\end{equation}
where $\hat {\textrm{H}}_j$ denotes a Hadamard gate on qubit $j$. Following preparation, we applied the single-spin field evolution represented by the unitary $\hat U_j(t)$ on spin $j$, and then appended the final analysis block $\hat D_2 = \prod_{j=1}^{5}\hat {\textrm{H}}_j$ for efficient measurement. The bulk magnetization was computed as $\Sigma_x=\sum_{j=2}^{4}\langle\hat{\sigma}_x^{(j)}\rangle$, and the string order parameter as $O_s=\langle\hat{\sigma}_x^{(1)}\hat{\sigma}_y^{(2)}\hat{\sigma}_z^{(3)}\hat{\sigma}_y^{(4)}\hat{\sigma}_x^{(5)}\rangle$.

For the cluster-Ising Hamiltonian experiment (results shown in Fig.~\ref{fig:Cluster-Ising}), the preparation block was
\begin{equation}
\hat D_1 = \prod_{j=1}^{11} e^{-i\tfrac{\pi}{4} \hat{\sigma}_x^{(j)}\hat{\sigma}_x^{(j+1)}}, 
\end{equation}
where $\hat R_n(\theta)=\exp{(-i\tfrac{\theta}{2}\mathbf{n}\cdot\boldsymbol{\hat{\sigma}})}$ denotes a single-qubit rotation of qubit $j$ about axis $\mathbf{n}$. The initial product state was prepared using \begin{equation}\hat D_0=\hat R_y^{(2)}\left(\frac{\pi}{2}\right)\hat R_y^{(11)}\left(\frac{\pi}{2}\right)\prod_{j\in E} \hat R_y^{(j)}(\pi),\end{equation} where $E$ is a randomly chosen subset of qubit indices for introducing bit flips to enable efficient sampling (see section ``Sampling advantage of digital-analog'').
The analog evolution implemented the transverse-field Ising model \begin{equation}
\hat H_\textrm{A}(\xi)=\sum_{i\neq j} J_{ij}\hat{\sigma}_x^{(i)}\hat{\sigma}_x^{(j)} +\xi g \sum_{j=2}^{11} \hat{\sigma}_z^{(j)},   
\end{equation} with the parameter $\xi=\pm1$.
To suppress dephasing errors from slow light-shift noise, we used a Hamiltonian engineering sequence \cite{morong2023engineering} \begin{equation} \hat U_{\textrm{A}}= \hat  R_x\left(\pi\right)e^{-\tfrac{i}{2}\hat H_{\textrm{A}}(\xi=-1)t}\hat R_x\left(\pi\right)e^{-\tfrac{i}{2}\hat H_{\textrm{A}}(\xi=1)t},\end{equation}
where \begin{equation}
\hat R_x(\theta)=\prod_{j=1}^{12}\hat R_x^{(j)}\left(\theta\right),\end{equation} 
which is equivalent to evolution by $e^{-i\hat H_{\textrm{A}}(\xi=1)t}$. The final analysis block 
\begin{equation}
\hat D_2 = \hat R_x^{(1)}\left(-\frac{\pi}{2}\right)\hat R_x^{(12)}\left(-\frac{\pi}{2}\right)
\end{equation}
enabled efficient measurement of both edge and bulk stabilizer correlators $G_j$ from readout of the state of qubit $j$.

For the four-body Hamiltonian evolution (data in Fig.~\ref{fig:Four_body}), the preparation sequence began with
\begin{equation}
\hat D_0 = \hat R_x^{(2)}\left(-\frac{\pi}{2}\right)\hat R_x^{(3)}\left(-\frac{\pi}{2}\right),
\end{equation} followed by
\begin{equation}
\hat D_1 =  \prod_{j=1}^{4}\hat {\textrm{H}}_j\prod_{j=1}^{4} e^{-i\tfrac{\pi}{4} \hat{\sigma}_x^{(j)}\hat{\sigma}_x^{(j+1)}}.
\end{equation}
The analog block consisted of evolution under
\begin{equation}
\hat H_\textrm{A}(\xi)=\sum_{i\neq j} J_{ij}\hat{\sigma}_x^{(i)}\hat{\sigma}_x^{(j)} +\xi g \sum_{j=1}^{4}\hat{\sigma}_z^{(j)}
\end{equation} for a total duration $t$. In Fig.~\ref{fig:Four_body}\textbf{b}, we used one of two analysis circuits $\hat D_2 = \hat {\textrm{H}}_1 \hat {\textrm{H}}_3$ or $\hat D_2 = \hat {\textrm{H}}_2 \hat {\textrm{H}}_4$. In Fig.~\ref{fig:Four_body}\textbf{c}, the analysis block was
\begin{equation}
\hat D_2 =\hat  R_x^{(1)}\left(-\frac{\pi}{2}\right)\prod_{j=1}^{2} e^{-i\tfrac{\pi}{8} \hat{\sigma}_x^{(j)}\hat{\sigma}_x^{(j+1)}}\left(\prod_{j=1}^{4}\hat {\textrm{H}}_j\right)\hat D_1^{\dagger} ,
\end{equation}
which includes non-Clifford gates. The observable \begin{equation}\Upsilon=\tfrac{1}{2}\langle\hat{\sigma}_x^{(2)}-\hat{\sigma}_z^{(1)}\hat{\sigma}_y^{(2)}-\hat{\sigma}_y^{(2)}\hat{\sigma}_z^{(3)}-\hat{\sigma}_z^{(1)}\hat{\sigma}_x^{(2)}\hat{\sigma}_z^{(3)}\rangle,\end{equation}
in Fig.~\ref{fig:Four_body} was then obtained directly from measurements on qubit 2. This can be seen by noting that $\Upsilon=\langle \hat{V}^{\dagger}\hat{\sigma}_z^{(2)}\hat{V}\rangle$ with $\hat V=\hat D_2\hat D_1$, using the identities
\begin{equation}
e^{-i \theta \hat{\sigma}_x^{(j)}\hat{\sigma}_x^{(j+1)}} \hat{\sigma}_z^{(j)} = \hat{\sigma}_z^{(j)} e^{+i \theta \hat{\sigma}_x^{(j)}\hat{\sigma}_x^{(j+1)}}
\end{equation}
and 
\begin{equation}
\hat{\sigma}_z^{(j)}\hat{\sigma}_x^{(j)} = i\hat{\sigma}_y^{(j)}.
\end{equation}
\section*{Numerical model}
In this Section, we present our numerical models for benchmarking the experiment.

We simulated the theoretical curves in Figs.~\ref{fig:Cluster-Field}-\ref{fig:Four_body} by emulating all digital circuits with unit fidelity. For the analog evolution, the Ising couplings were calculated as \begin{equation}J_{i, j} = \sum_k \frac{\eta_{i, k}\eta_{j, k} \Omega_i \Omega_j}{2(\mu+\omega_1-\omega_k)}, \end{equation}
where $\eta_{i, k}$ is the Lamb-Dicke parameter for spin $i$ and phonon mode $k$, defined by the non co-propagating Raman wavevector. The single-ion Rabi frequencies ${\Omega_i}$ were set from relative Rabi calibrations and the experimentally measured average nearest-neighbor coupling $J=\tfrac{1}{11}\sum_{j} J_{j, j+1}$, and $\omega_k$ are the phonon mode frequencies. 

For Figs.~\ref{fig:Cluster-Field} and \ref{fig:Four_body}, the simulations assume no physical errors. In Fig.~\ref{fig:Four_body} we take $J=2\pi \times 200$ Hz and next-nearest-neighbor strength $B=2\pi \times 78$ Hz. 

For the dynamics in Fig.~\ref{fig:Cluster-Ising}, we include ion-dependent dephasing that is calibrated independently by measuring the decay of Ramsey fringes of individual spins in our 15-ion chain. The Ramsey measurement is performed in the presence of a far-off-resonant Raman beam detuned by  $-2\pi\times 1$ MHz from the lowest-frequency radial mode, which suppresses spin-spin interactions while preserving the same light-shift noise. Following Refs.~\cite{schuckert2025observation, miao_probing_2025}, the simulated evolution is divided into time steps of duration $\Delta t = 20$ $\mu$s. After each noiseless Hamiltonian step, a phase-flip gate $\hat R_z^{(i)}(\pi)$ is applied to each spin $i$ with probability $p_i=\frac{1}{2}\gamma_i \Delta t$, where $\gamma_i$ are the experimentally measured single-spin Ramsey contrast decay rates [14(7), 39(3), 55(6), 44(2), 46(3), 60(3), 49(3), 51(7), 42(12), 36(5), 40(13), 24(8)] $\mu$s$^{-1}$.
This procedure reproduces the observed dephasing without introducing any free parameters; all simulation inputs are taken directly from experimental calibration.

\begin{figure*}[htbp]
\begin{centering}
 \includegraphics[width=6in]{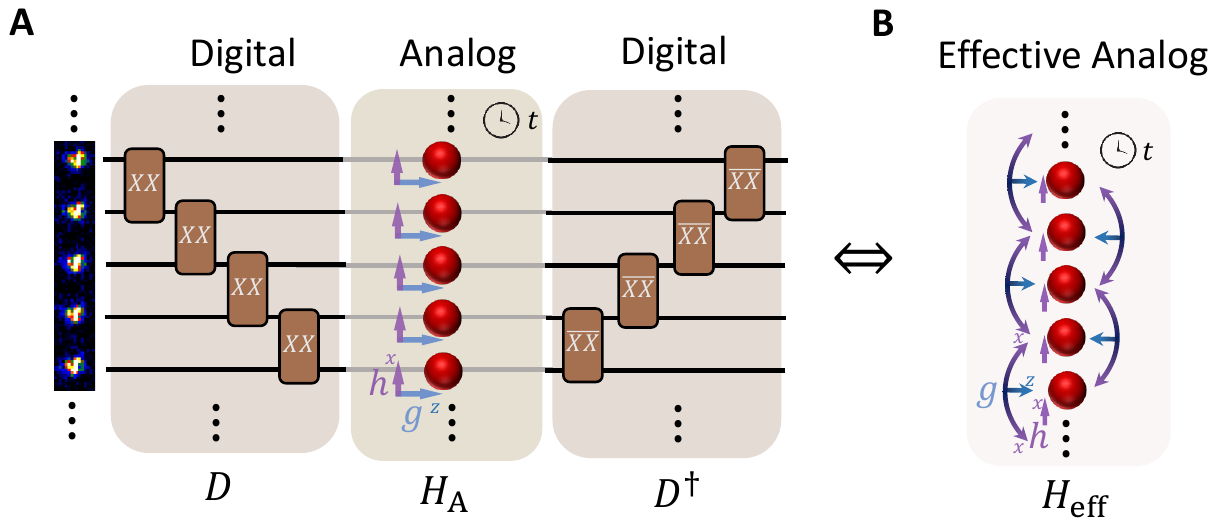}
\par\end{centering}
\centering{}\caption{\textbf{Cluster-Field Hamiltonian}. \textbf{a}, An analog Hamiltonian composed of one-body terms is interleaved with digital layers containing two qubit gates between qubits $i$ and $j$ of the form $XX=\exp(-i\tfrac{\pi}{4}\hat{\sigma}_{x}^{(i)}\hat{\sigma}_{x}^{(j)})$ and $\overline{XX}=(XX)^{\dagger}$. Fluorescence imaging (left) identifies the middle five physical ions corresponding to each qubit/spin in a 15-ion chain. \textbf{b}, The resulting effective Hamiltonian contains one- and emergent three-body interactions not present in the original analog Hamiltonian. \label{fig:Intro}}
\end{figure*}
\stepcounter{extended_data_fig}

\begin{figure*}[htbp]
\begin{centering}
 \includegraphics[width=6in]{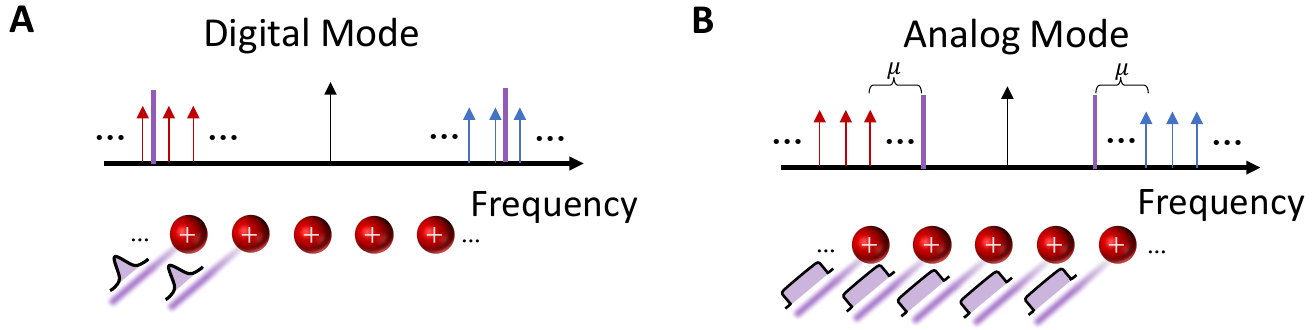}
\par\end{centering}
\centering{}\caption{\textbf{Physical realizations of digital and analog primitives in trapped-ion systems}. \textbf{a}, Digital two-qubit operations transiently couple select spin pairs to the joint motion of the ions via amplitude-modulated drive near the motional sidebands. \textbf{b}, Analog interactions are generated by a continuous spin-dependent force that is off-resonant from motional sidebands and induces weak, dispersive phonon-mediated couplings between all spins. \label{fig:Exp_ions_digital_analog}}
\end{figure*}
\stepcounter{extended_data_fig}

\begin{figure}[htbp]
\begin{centering}
 \includegraphics[width=8.6cm]{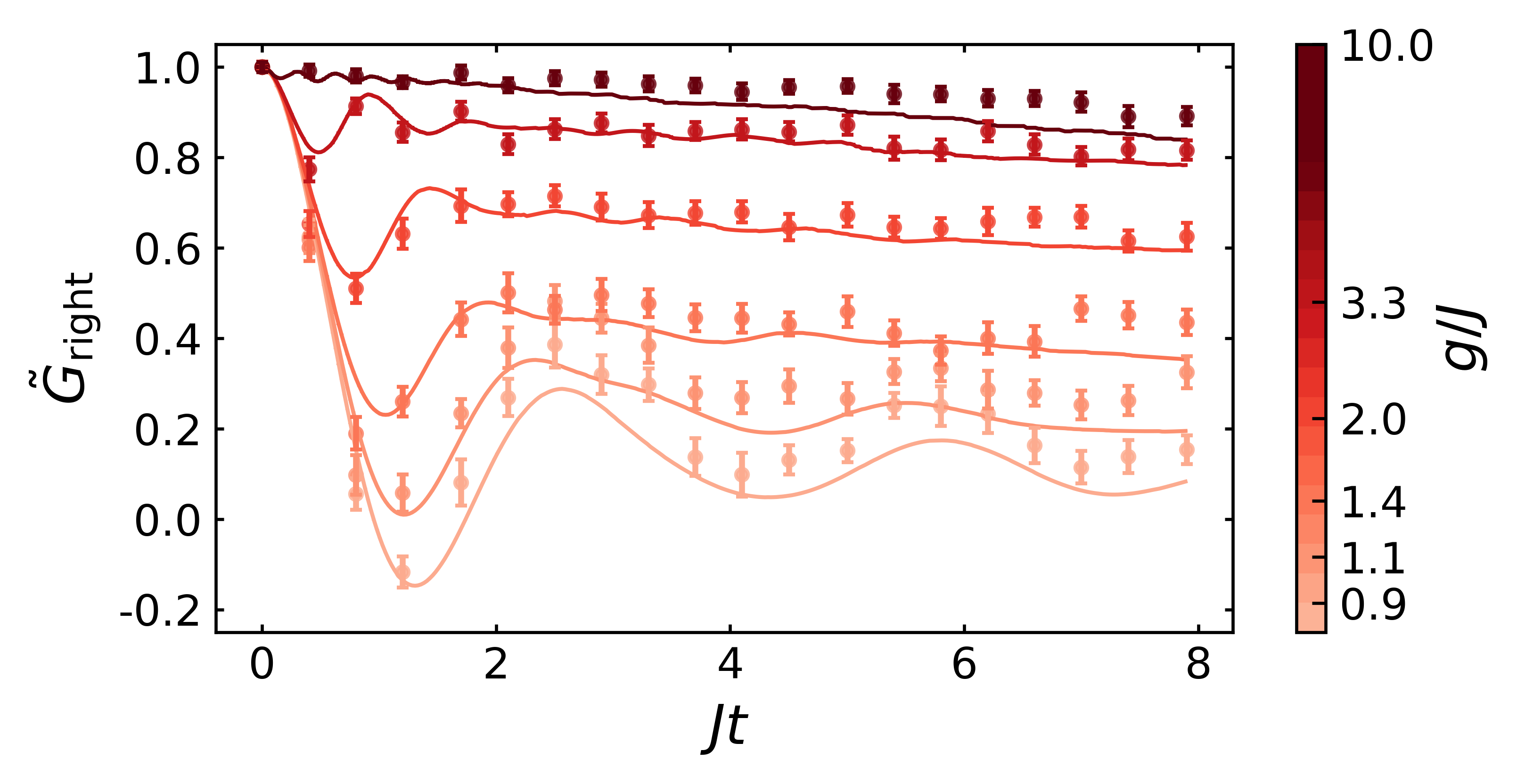}
\par\end{centering}
\centering{}\caption{ Time evolution of the normalized edge correlator $\tilde G_{\textrm{right}}(t)$ at infinite temperature for varying interaction ratios $g/J$. The observed long-lived correlations reveal prethermal persistence despite moderate Ising perturbation, and are similar to the behavior of the left edge (See Fig.~\ref{fig:Cluster-Ising}\textbf{c}). Symbols denote measured data, and lines show numerical simulations. Error bars denote the quadrature sum of the $1\sigma$ binomial uncertainties of each state and the $1\sigma$ standard error of the mean over all input states.
\label{fig:Right_edge_correlator}}
\end{figure}
\stepcounter{extended_data_fig}

\begin{figure*}
    \centering
    \includegraphics[width=6.8in]{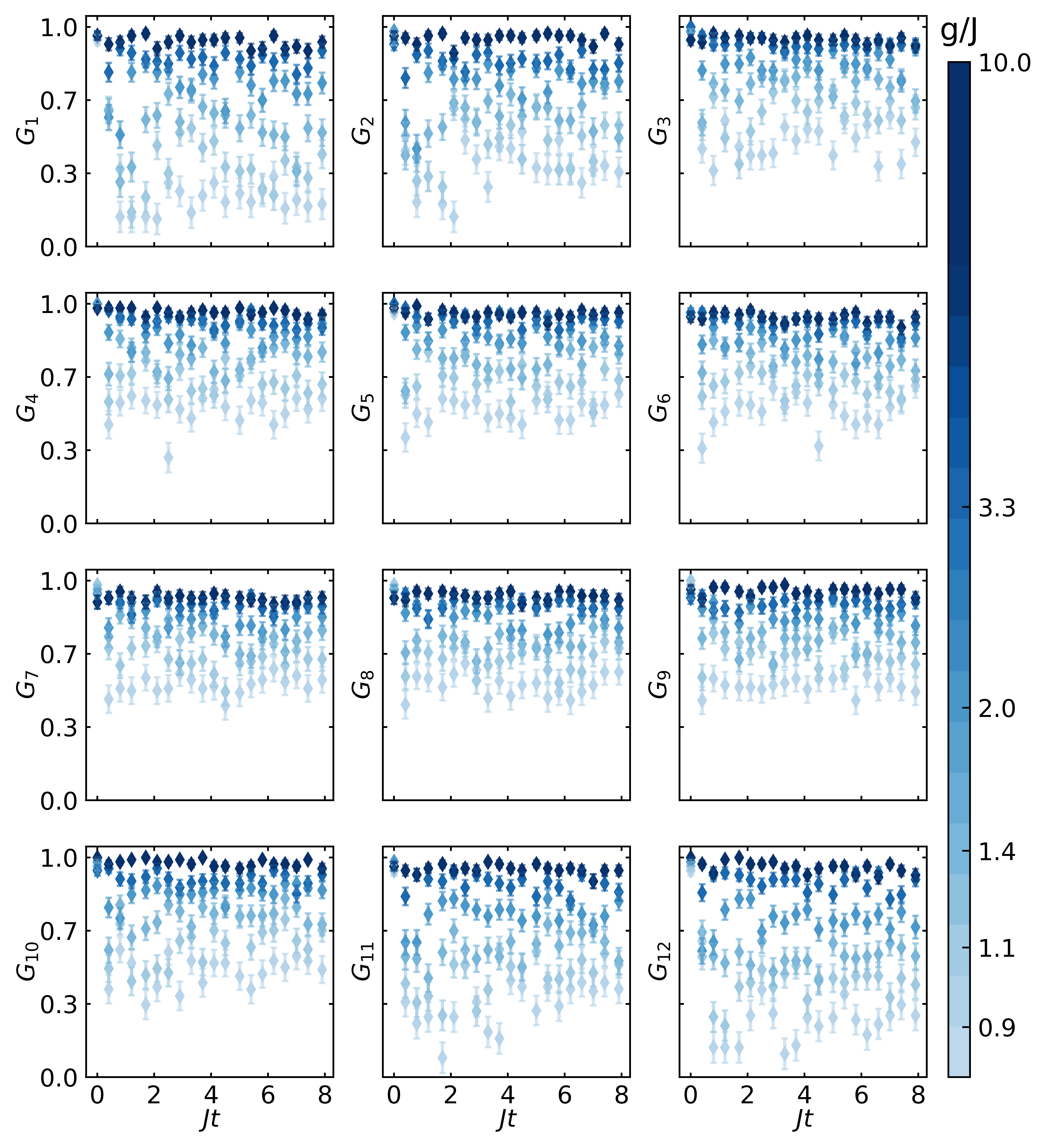}
    \caption{\textbf{Site-resolved correlator dynamics following an Ising quench from the cluster ground state.} Time evolution of the unnormalized stabilizer correlators $G_j(t)=\langle \hat{S}_j(t)\rangle\langle \hat{S}_j(0)\rangle$ with the stabilizer operator $\hat{S}_j=\hat{\sigma}_x^{(j-1)}\hat{\sigma}_z^{(j)}\hat{\sigma}_x^{(j+1)}$ in the bulk ($2\leq j\leq 11$) and  $\hat{S}_1=\hat{\sigma}_z^{(1)}\hat{\sigma}_x^{(2)} $, $\hat{S}_L=\hat{\sigma}_x^{(L-1)}\hat{\sigma}_z^{(L)}$ at the edges of (associated with an open boundary chain spin). At $t=0$, the Ising interaction is turned on. Colors indicate the relative $g/J$ strength, local stabilizers preserving their values despite moderate Ising interaction. Symbols denote measured data; error bars indicate $1\sigma$ binomial uncertainties
    \label{fig:gnd_state_correlator}}
\end{figure*}
\stepcounter{extended_data_fig}

\begin{figure*}
    \centering
    \includegraphics[width=6.8in]{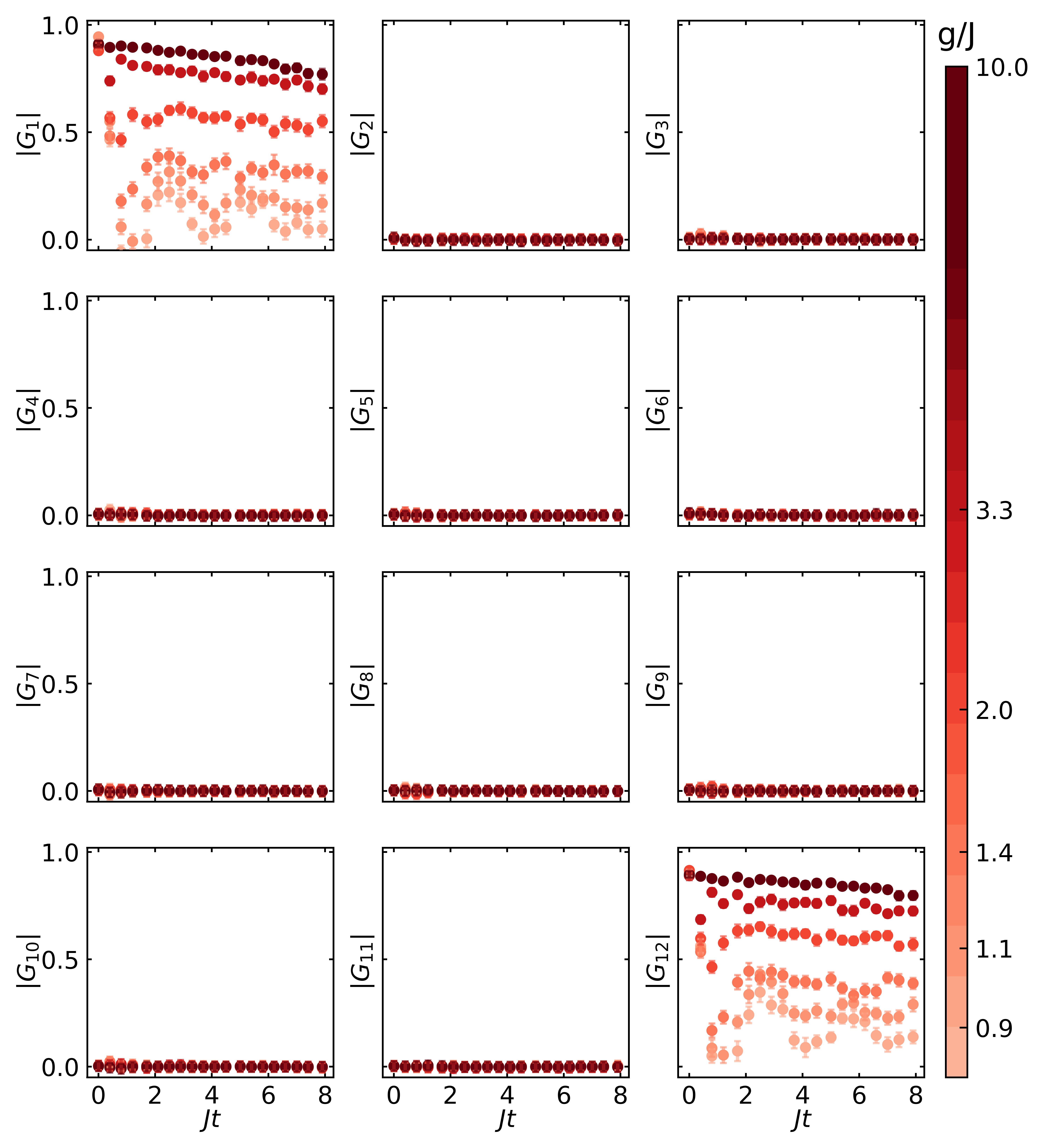}
    \caption{\textbf{Site-resolved correlator dynamics following an Ising quench at infinite temperature.} Time evolution of the absolute value of the unnormalized stabilizer correlators $G_j(t)=\langle \hat{S}_j(t)\rangle\langle \hat{S}_j(0)\rangle$ averaged over all sampled states with the stabilizer operator $\hat{S}_j=\hat{\sigma}_x^{(j-1)}\hat{\sigma}_z^{(j)}\hat{\sigma}_x^{(j+1)}$ in the bulk ($2\leq j\leq 11$) and  $\hat{S}_1=\hat{\sigma}_z^{(1)}\hat{\sigma}_x^{(2)} $, $\hat{S}_L=\hat{\sigma}_x^{(L-1)}\hat{\sigma}_z^{(L)}$ at the edges of (associated with an open boundary chain spin). At $t=0$, the Ising interaction is turned on. Colors indicate the relative $g/J$ strength, local stabilizers in the bulk are close to zero while the edge correlators stabilize to their prethermal values for moderate Ising interaction. Symbols denote measured data; error bars indicate $1\sigma$ binomial uncertainties
    \label{fig:infintie_temp_state_correlator}}
\end{figure*}
\stepcounter{extended_data_fig}

\begin{figure*}[htbp]
\begin{centering}
 \includegraphics[width=5in]{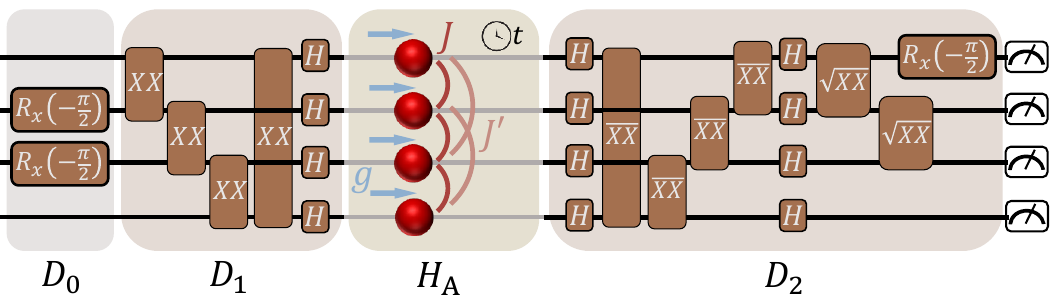}
\par\end{centering}
\centering{}\caption{\textbf{Digital-analog protocol for the dynamics in Fig.~\ref{fig:Four_body}}. The circuit begins with the preparation block $\hat D_0$, followed by the entangling layer $\hat D_1$, and analog evolution under the native TFIM Hamiltonian $H_{\textrm{A}}$ (similar to Fig.~\ref{fig:Cluster-Ising}). Analysis circuits $\hat D_2$ enable measurement of the observables shown in Fig.~\ref{fig:Four_body}\textbf{c}, with additional configurations detailed in Methods.  Note that  $\hat D_0$, $\hat D_1$, $\hat D_2$ differ from Fig.~\ref{fig:Cluster-Ising}, so the effective Ising interaction is now along $z$, while the transverse field is along $x$. $H$ denotes Hadamard gare,  $XX=\exp(-i\tfrac{\pi}{4}\hat{\sigma}_{x}^{(i)}\hat{\sigma}_{x}^{(j)})$,  $\overline{XX}=(XX)^{\dagger}$ and $\sqrt{XX}=\exp(-i\tfrac{\pi}{8}\hat{\sigma}_{x}^{(i)}\hat{\sigma}_{x}^{(j)})$ denote two-qubit gates used between qubits $i$ and $j$. \label{fig:Four_body_protocol}}
\end{figure*}
\stepcounter{extended_data_fig}

\stepcounter{extended_data_fig}

\stepcounter{extended_data_fig}

\begin{acknowledgments} 
This work was funded by the NSF STAQ project (Phy-2325080). Support is also acknowledged from the U.S. Department of Energy, Office of Science, National Quantum Information Science Research Centers, Quantum Systems Accelerator. O.K. acknowledges support from the National Science Foundation Division of Physics (Investigator-Initiated Research Projects) under Award No. PHY-2512966. 
A.V.G.~was also supported in part by the DoE ASCR Quantum Testbed Pathfinder program (awards No.~DE-SC0019040 and No.~DE-SC0024220), NSF QLCI (award No.~OMA-2120757), ARL (W911NF-24-2-0107), AFOSR MURI, ONR MURI, and NQVL:QSTD:Pilot:FTL. A.V.G.~also acknowledges support from the U.S.~Department of Energy, Office of Science, Accelerated Research in Quantum Computing, Fundamental Algorithmic Research toward Quantum Utility (FAR-Qu). Specific
product citations are for the purpose of clarification only and
are not an endorsement by the authors or NIST.

\end{acknowledgments}
\clearpage
\bibliography{Refs}

\end{document}